\documentclass{article}
\usepackage{spconf,amsmath,graphicx, color}
\usepackage{enumitem}

\def \etal {et al.~}

\title{Detecting expressions with multimodal transformers}
\name{Srinivas Parthasarathy, Shiva Sundaram}
\address{Amazon}
\begin{document}
	
	\maketitle
	\begin{abstract}
		Developing machine learning algorithms to understand person-to-person engagement can result in natural user experiences for communal devices such as Amazon Alexa. Among other cues such as voice activity and gaze, a person’s audio-visual expression that includes tone of the voice and facial expression serves as an implicit signal of engagement between parties in a dialog. This study investigates deep-learning algorithms for audio-visual detection {\color{black}of} user’s expression. We first implement an audio-visual baseline model with recurrent layers that shows competitive results compared to current state of the art. Next, we propose the transformer architecture with encoder layers that better integrate audio-visual features for expressions tracking. Performance on the Aff-Wild2 database shows that the proposed methods perform better than baseline architecture with recurrent layers with absolute gains approximately 2\% for arousal and valence descriptors. Further, multimodal architectures show significant improvements over models trained on single modalities with gains of up to 3.6\%. Ablation studies show the significance of the visual modality for the expression detection on the Aff-Wild2 database.
		
	\end{abstract}
	
	\section{Introduction}
	
	Among the different forms of human interaction, audio-visual communication is principal. Humans inherently process and react to their social surroundings which manifests as an exterior display of expression, forming a vital part of this communication. 
	With the growth of communal devices that users can talk and communicate with around their homes, detecting and tracking expressions that understand users' affect in addition to spoken words can help build better natural interfaces. Such devices play important roles in various fields such as communication, health care \cite{Cummins_2015}, security \cite{Clavel_2008}, and education \cite{Litman_2004}.
	
		\begin{figure}[tb]
		\centering
		\includegraphics[width=0.6\linewidth]{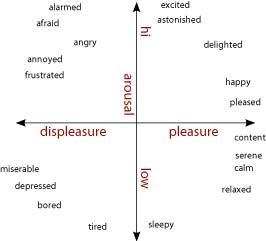}
		\caption{Mapping between attribute descriptors and categorical labels for expressions, Russell \cite{Russell_1980}}
		\label{fig:labels}
	\end{figure}
	
	In addition to speech \cite{el2011survey}, humans naturally use non-verbal modalities such as pose, gestures and facial cues for communicating with affect. Therefore recognition algorithms have focused on explicit forms of expressions including a change in the tone and style of voice \cite{Cowie_2003}, facial expressions \cite{Sariyanidi_2015, kollias2017recognition}, body language and posture \cite{Gunes_2006, Coulson_2004} as well a fluctuation in physiological signals such as heart rate \cite{Martinez_2013, Kim_2004}. However, tone of voice (audio) and facial cues (video) remain primary forms of communicating affect \cite{Mehrabian_2007}. Also, the proliferation of audio and video sensors in most devices make these appealing for expression detection. 
	Expression \footnote{Affect is broadly categorized as emotions, expressions and sentiment by different communities. Since we deal with audio and visual cues we use the terms emotion and expression interchangeably while referring to affect.} detection can be predominantly divided into two methods based on the type of labels. A discrete classification approach, categorizes the continuum of expressions into few categories such as happiness, sadness, anger and neutral \cite{Cowie_2003} and a descriptor based approach measures expressions in predominantly two dimensions: arousal (level of excitement) and valence (negative to positive levels) \cite{Russell_1980}. 
	{\color{black} 
		The circumplex model, discussed by Russell, \cite{Russell_1980} (Fig. \ref{fig:labels}) remains a popular technique for mapping emotional categories to arousal and valence values.} 
	Recently, there is also an increased focus to predicting attribute values as they can be used to describe expressions on a finer granularity \cite{Parthasarathy_2017_3}, for capturing nuanced emotions. 

This work is about the multi-modal detection of arousal and valence attributes, with a focus on the visual and audio modalities. We implement a transformer model \cite{Vaswani_2017attention} with a cross-modal attention layer to fuse the auditory and visual features. The proposed model attends to significant cues within each modality through self-attention encoder layers as well as across modality through the cross-modal attention layer, capturing temporal context within audio-visual clips. We evaluate the proposed model on the Aff-Wild2 database \cite{kollias2018aff} for detecting arousal and valence scores. We first establish a competitive baseline that uses recurrent layers to learn the temporal context in the audio and video signals. We show that this baseline model surpasses other published results learned on single modalities. Next, we show that the proposed audio-visual transformer performs significantly better than the baseline architecture, with absolute gains of 1.7\% for valence and 1.9\% for arousal over our baseline results. This indicates the efficacy of the model architecture for the expression detection task; producing state of the art performance on Aff-Wild2 dataset. Further, the proposed models competitively perform against other contemporary models, producing state of the art performance. Finally, we also study the impact of each modality on the proposed model through an ablation study indicating that the video modality might be dominant for recognizing expressions.

	This paper contributes the following - a) A competitive recurrent neural network (RNN) baseline model for modeling expression detection task for single and multiple modalities b) A multi-modal transformer architecture with cross-modal attention layers to fuse the audio-visual modalities, that produces state of the art performance.

	\section{Background}

	\subsection{Audio Cues}
	A number of research efforts have studied affect recognition from audio cues. Traditional methods focus on both categorical recognition over discrete utterances \cite{Lee_2004} as well as continuous prediction \cite{Fontaine_2007}. A variety of features are generally used to recognize emotions. These include log-spectrograms, mel-frequency filterbanks \cite{Aldeneh_2017}. Several researches have also studied para-linguistic features for emotion recognition \cite{Schuller_2009, Schuller_2010}. These features are generally computed over short overlapping frames that descriptive of the changes in the vocal effort due to changes in affective expression. Zeng \etal provides more details on architectures for audio expression detection \cite{Zeng_2009}.
	
	\subsection{Facial Expressions}
	Several works have focused on detecting facial expressions with visual cues. Facial expression is mainly divided into facial emotions or facial action units (facial muscle movements) \cite{Zeng_2009}. Detection methods also vary by task: static recognition on images \cite{shan2009facial, liu2014facial} or continuous prediction that look at sequences of images \cite{jung2015joint}. Traditional methods use a face-alignment technique to crop the face from the background. Further, geometric visual cues are generally used for the recognizing facial expressions. These cues range from shape of facial features (shape of eyes, mouth, eyebrows) and location of facial features (corner of the eyes, corner of lips) \cite{chang2006manifold, pantic2006dynamics}. A landmark detector is generally used to extract these visual cues \cite{mollahosseini2016going}. Recently, with the growth of deep learning techniques there is an increased focus on expression detection in the wild with data-driven models \cite{mollahosseini2016going, mollahosseini2017affectnet}. {\color{black}Similar to these previous studies, our work also employs a face detector} to extract face crops and further process facial features using a pre-trained model (Section \ref{sssec:video})

	\subsection{Audio  Visual Expression Detection}
	\label{ssec:av_detection}
	{\color{black}A number of studies have also developed  machine learning models for  predicting  audiovisual expression \cite{albanie2018emotion, ouyang2017audio, vielzeuf2017temporal}. They have shown that audio and visual cues provide complementary information for expression detection. While audio generally provides cues for prediction of arousal with change in tone correlating to change in arousal, video cues support the prediction of valence. Audio and video modalities are either combined at feature-level (early fusion), at decision level (late fusion) or a hybrid version \cite{Zeng_2009}. Deep learning techniques have given rise to early fusion of the multimodal features with data driven approaches \cite{tzirakis2017end}. This work adds to these previous studies by proposing an attention mechanism to fuse latent audio-visual cues for expression detection.

	\subsection{Attention for expression}
	 Broadly, the  attention mechanism can be considered the allocation of resources to pertinent information in the signal. For audio cues, Chen \etal \cite{chen20183} proposed a 3-D convolutional recurrent neural network with an attention mechanism to focus on detected speech frames. Their approach shows improvement in the unweighted recall of the emotion classification. For facial cues, Xiaohua \etal \cite{xiaohua2019two} propose a two stage attention network to model the relationship between different positional features on the face. Hazarika \etal  \cite{hazarika2018conversational} employ a soft attention mechanism for capturing memory in audio, visual and textual cues for dyadic interactions. Conventionally, studies focused solely on recurrent layers for capturing the temporal context to predict expressions. Mirsamadi \etal \cite{mirsamadi2017automatic} include a local attention mechanism to pool emotion specific audio cues on top of recurrent layers. There is also growing concentration on studying self-attention mechanisms to replace recurrent neural networks for expression recognition. Li \etal \cite{li2019improved} propose a self-attention module on top of convolutional and recurrent layers to improve emotion recognition performance from speech. Rahman \etal \cite{rahman-etal-2020-integrating} propose methods to integrate audio and visual cues to transformers pretrained on text cues. Their work uses audio-visual cues to gate the text features for emotion recognition. In contrast to the studies outlined above, this work uses a novel cross-modal attention mechanism to fuse audio-visual cues to detect expressions.

	\section{Proposed Methodology}
		\begin{figure}[tb]
		\centering
		\includegraphics[width=\linewidth]{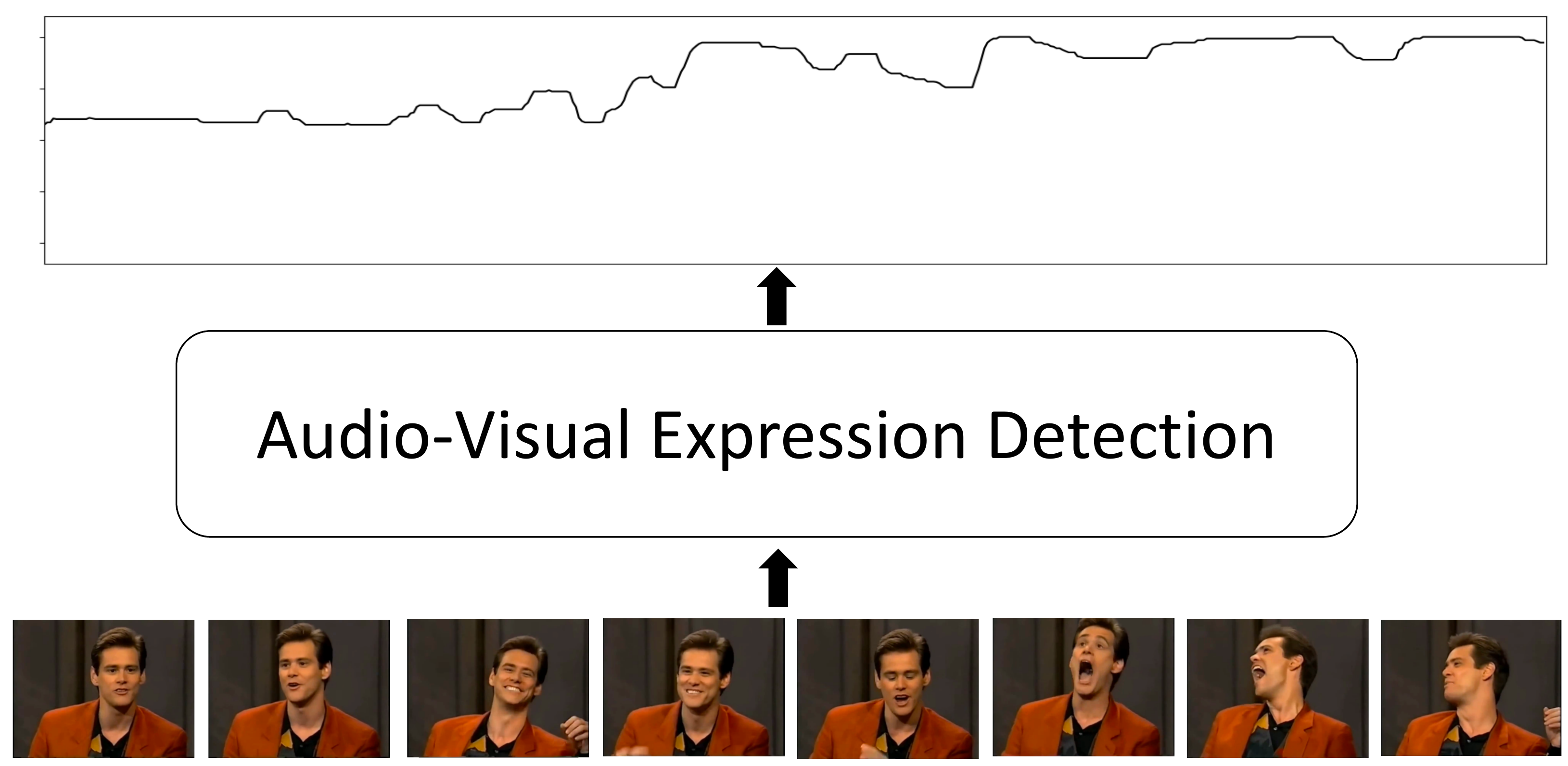}
		\caption{{\color{black} Continuous expression recognition. The system takes audio-visual frames as input and predicts arousal valence value for each frame. }}
		\label{fig:overall}
	\end{figure}

	This work addresses the continuous recognition of expression (Figure \ref{fig:overall}). Given a sequence of audio-visual frames, the task is to predict the arousal and valence values for each frame in the sequence. 

	As outlined in Section \ref{ssec:av_detection}, while audio and video streams can be fused in various ways, an intuitive approach would be to let the model learn the integration based on the data. The attention mechanism, provides an appealing framework to learn this fusion. While there are several methods to learn attention, the dot-product attention, discussed by Vaswani \etal \cite{Vaswani_2017attention} has shown state of the art results for various tasks \cite{devlin2018bert}. We propose a cross-modal attention with dot-product attentions \cite{Vaswani_2017attention}, to learn feature representations that are beneficial for overall prediction (Figure \ref{fig:architecture}). Cross-modal attention has been successfully used for audio-visual automatic speech recognition \cite{paraskevopoulos-etal-2020-multimodal}. We believe the features learned with a cross-modal attention layer will be invariant to occlusions or the absence of data in either modality. For e.g when the users face is not in the field of view (FoV) of the camera the model will focus on audio cues to detect the users expression and vice versa when the user is not speaking \cite{parthasarathy2020training}. 
	
	The overall architecture consists of two encoders, one each for audio and the video modality. Each encoder consists of multiple self-attention layers. Each self-attention layer works by constructing three matrices $K$, $V$ and $Q$ from the previous layer outputs, where $K$ and $V$ act as keys and values in a dictionary and $Q$ acts a query to compute the attention. The output values are then given by
	\vspace{-0.15cm}
	\begin{equation}
	Y = \sigma(KQ^T)V
	\label{eq:att}
	\vspace{-0.15cm}
	\end{equation}
	where $\sigma$ denotes the softmax operation, computing the attention as a dot product between the $K$ and $Q$ matrices. 
	
	With the self-attention layers, each encoder individually attends to significant cues for the corresponding modality. The encoder embeddings are then fused using two cross-modal attention layers. The cross-modal attention layers also consist dot-product attentions similar to the encoder layers. with the key difference being $K$ and $V$ matrices are computed from one modality and the $Q$ matrix is computed from the opposite modality (Figure \ref{fig:architecture}). Through the cross modal attention layers, the model learns to attend to cues from the other modality. The cues from the respective branches are then fused using a weighted sum of their values using simple scalar weights $\alpha$, $\beta$ . The combined features are then used to predict the arousal and valence values.
	\begin{figure}[tb]
		\centering
		\includegraphics[width=\linewidth]{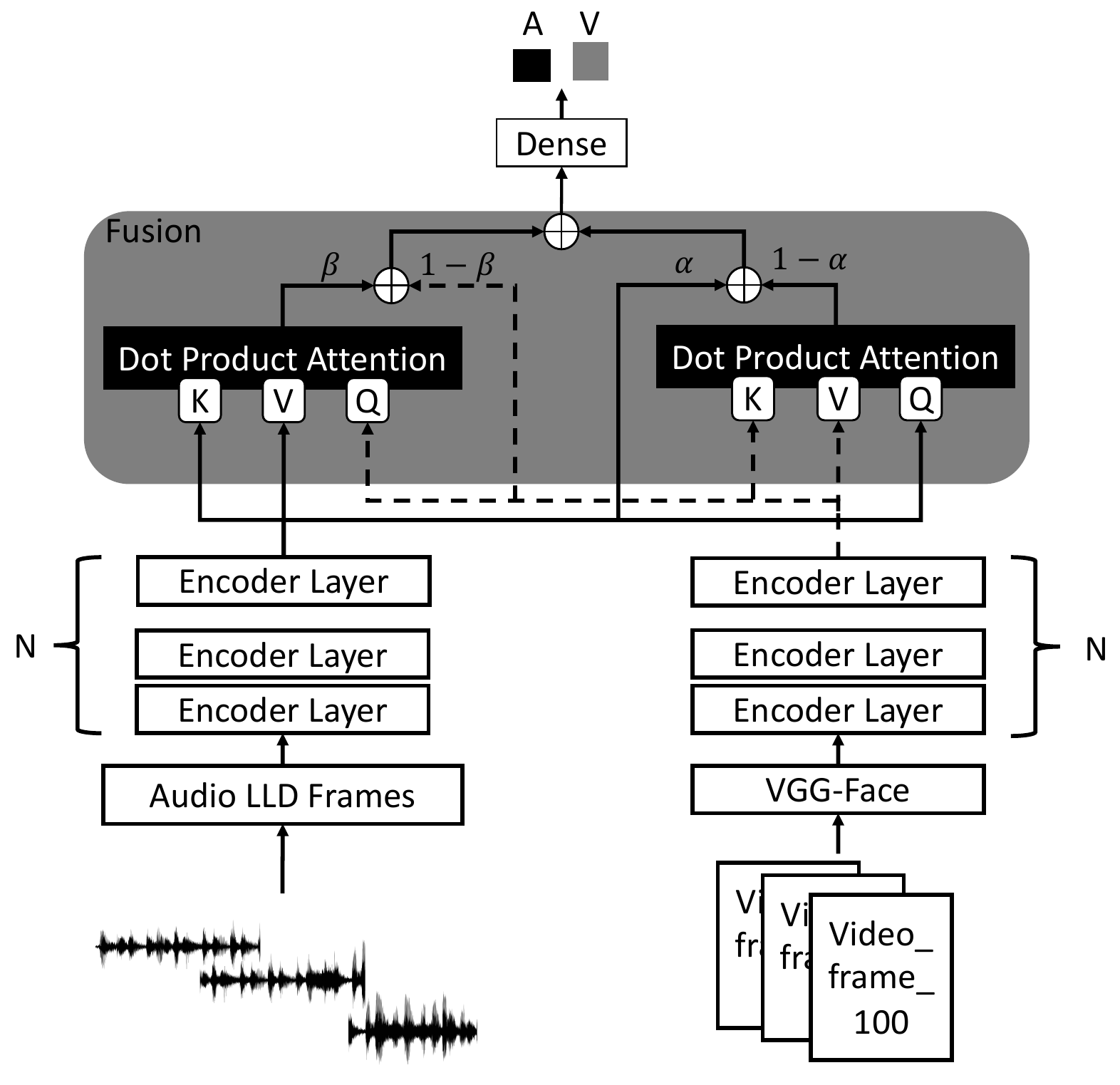}
		\caption{Proposed system architecture. A cross-modal scaled dot-product attention layer is used to project the visual data into the audio feature space and vice versa, followed by an additive fusion.}
		\label{fig:architecture}
	\end{figure}
	\section{Experimental setup}
	
	\subsection{Dataset}
	This study uses the Aff-Wild2 database \cite{kollias2018aff}. While other datasets exist with carefully curated data, the Aff-Wild2 data is unique for its in the wild recordings. This is significant as data is truly recorded in various conditions, for e.g subject could be silent while still providing facial expression like a smile. These conditions make it ideal to study the fusion of audio and video cues, an important aspect of this work. The dataset contains 558 videos of mostly single persons expressing themselves. Videos are captured at a frame rate of 30 \emph{frames per second} (fps) with around 2.8 million video frames in the dataset. The dataset is annotated for 3 main behavior tasks (basic expression classification, valence-arousal estimation and action unit detection.) This paper focuses on frame-level valence-arousal estimation which the dataset provides. Every frame in the dataset is annotated for valence and arousal values in a continuous scale between [-1, 1] by 4 or 8 annotators. Further the dataset is divided into speaker independent train, validation and test subsets consisting of 350, 70 and 138 videos respectively.
	
	\subsection{Features and Preprocessing}
	\label{ssec:features}
	\subsubsection {Video}
	\label{sssec:video}
	We follow previous works to extract bounding boxes to detect frontal faces on the video frames \cite{kollias2019expression}. We use an SSD based detector with a ResNet architecture to detect faces in every frame \cite{Liu_2016}. The image is cropped based on the bounding box and resized to a dimension of 96x96x3. All pixel intensities are scaled between [-1, 1].
	\subsubsection {Audio}
	For the audio cues we extract the features introduced for the para-linguistic challenge at Interspeech 2013 \cite{Schuller_2013}. These features have shown state of the art performance for various emotion recognition tasks. The feature set consists of \emph{low-level descriptors} (LLD) that are extracted over 25ms windows with a 10ms step and further statistical functions that are calculated on top of the LLDs to get segment level features \emph{high level descriptors} (HLD). While the use of HLD is more prevalent, the high dimensionality of the overall features (6373) is a particular drawback. Therefore we avoid this and rely only on the 65 LLDs which consists of spectral features such as the \emph{Mel Frequency Cepstral Coeffecients} (MFCC), energy features such as loudness and voice based features such as fundamental frequency (F0). Features are Z-normalized based on the train set statistics.
	
	\subsection{Architecture}
	We follow the transformer architecture proposed in \cite{Vaswani_2017attention}. First two sets of encoder layers process the audio and video inputs respectively. Each encoder layer uses performs a self-attention with a multi head attention  (Eq. \ref{eq:att}). We use 512 nodes in each encoder layer with four attention heads. The fusion layer consists of two dot-product attention layers with 512 nodes each. These attention layers attend to the other modality. Embeddings from the cross-modal attention layers are linearly combined using scalar trainable parameters $\alpha$, $\beta$.  Predictions are made with a dense output layer.
	
	Predictions are made on sequences of length 100, corresponding to roughly 3s of video (Figure \ref{fig:overall}). Besides extracting the face crops, the video frames are passed through a pretrained VGGFace network \cite{Parkhi15} to extract descriptors from the face region in each frame. We feed the first fully connected layer of the network containing 4096 nodes as input to the transformer video encoder. We perform a synchronous fusion of audio and video features. To synchronize the audio and video, audio frames are downsampled by a factor of 3.3. Further we concatenate the audio LLDs over a two second window to match the feature dimension (4096) of the video. These concatenated features are fed to the audio encoder of the transformer. We use the \emph{concordance correlation coefficient} (CCC), which calculates loss at the sequence level, as both the loss (1 - CCC) and evaluation metric.
	
	\begin{equation}
	\rho_c = \frac{2 \rho \sigma_x \sigma_y}{\sigma_{x}^2 + \sigma_{y}^2 + (\mu_x - \mu_y)^2}
	\end{equation}
	{\color{black}where $\rho$ corresponds to the Pearson's correlation, $\sigma$ corresponds to the standard deviation, $\mu$ corresponds to the mean and $x,y$ correspond to the ground-truth and predictions.} The model is trained using an Adam optimizer with an initial learning rate of $1e^{-5}$. The best model is chosen over the validation set
	
	\subsection{Baseline models}
	\label{ssec:baseline}
	We compare proposed models directly against pretrained baselines from previous studies as well as modifications to these models. We establish baselines for the audio (A) and video (V) modalities independently as well as audio-visual (A+V) baselines. Other than the pretrained models, all models are trained using the CCC loss.
	{\color{black}
	\begin{itemize}[leftmargin=0pt]
		\item[] \textbf{AffWildNet + Static (V)} Kollias \etal \cite{kollias2019deep} models are pretrained on the the Aff-Wild database \cite{kollias2019deep}, a predecessor of the Aff-Wild2 database, on the video modality only. The static models make predictions on each image independently using the VGGFace architecture which consists of convolutional layers, one fully connected (FC) along with an output layer for the predictions. We try two dimensions for the FC layer - 2000, 4096. Note that the proposed models are built on top of the VGGFace architecture used in this baseline and therefore share preprocessing steps.
		\item[] \textbf{AffWildNet + Dynamic (V)} Kollias \etal \cite{kollias2019deep} is pretrained on the Aff-wild database. The dynamic model on the other hand uses a sequence of images. We use the same sequence length of 100 used for the proposed transformers. The dynamic model contains all the layers in the static model (with FC layer dimesion 4096) plus two recurrent layers with \emph{gated recurrent units} (GRU). The GRU layers contain 128 nodes each.
		\item[] \textbf{RNN (A)} is an audio baseline model that uses two GRU layers on top of the LLD features.
		\item[] \textbf{VGGFace-RNN (V)} has the same architecture as \textbf{AffWildNet + Dynamic (V)} but is additionally trained on Aff-Wild2.
		\item[] \textbf{VGGFace-RNN (A + V)} is an audio-visual baseline model that concatenates the output from the recurrent layers of the audio only \textbf{RNN (A)}) and video only (\textbf{VGGFace-RNN (V)}) model. This is followed by two additional GRU layers and dense layer for output predictions. The proposed models are mainly compared against this audio-visual baseline.
		\item[] \textbf{NISL (V)} Deng \etal \cite{deng2020multitask} pretrained CNN + RNN models trained using a multitask framework for emotional categories, facial action units and emotional attributes. Authors also use external datasets to balance the distribution of emotions in the Aff-Wild2 dataset. 
		\item[] \textbf{$M^3$T (A+V)} Zhang \etal \cite{zhang2020m3t} pretrained models using a 3D convolutional network and bidirectional recurrent neural network. The model is trained with a multitask setup with joint prediction emotion categories helping the prediction of valence attribute. While \cite{zhang2020m3t}, used an attention mechanism, on top of the recurrent layers, to fuse the audio and video features, best results are obtained by simple concatenation. {\color{black} Zhang et al. \cite{zhang2020m3t} note that their attention model may not have reached full model convergence compared to model trained with concatenation. Note that our preliminary experiments on a cross-modal layer showed benefits over simple concatenation of the features from the two encoder branches in the proposed model.}
		\end{itemize}
	Note that NISL and $M^3T$ are {\color{black}external studies} that report state of the art results on the Aff-Wild2 database. {\color{black} The VGGFace-RNN (A + V) is our proposed baseline that is thoroughly trained on the Aff-Wild2 database and comparable to the external studies} While the state of the art models share aspects such as convolutional layers to process video features, the biggest difference with our proposed work is the use of dot-product attentions. We also analyze the impact of individual modality in training the multimodal architecture (Section \ref{ssec:ablation}).
 }
	\section{Results}
	
	\begin{table}[t]
		\caption{Results on the Aff-Wild2 database. Significance analysis with Fishers z-transformation ($p<0.01$). \textbf{Ar} corresponds to Arousal and \textbf{Va} corresponds to Valence}
		\label{tab:baseline_results}
		\centering
		\begin{tabular}{c|c|c}
			\hline
			\hline
			\textbf{Model}  & \textbf{CCC-Va} &\textbf{CCC-Ar}\\
			\hline
			\hline
			AffWildNet + Static (V) (2000) \cite{kollias2019deep}  & 0.176   & 0.198 \\
			AffWildNet + Static (V) (4096) \cite{kollias2019deep}  & 0.244  & 0.297  \\
			AffWildNet + Dynamic (V)  & 0.315  & 0.357 \\
			RNN (A) & 0.105 & 0.168 \\
			VGGFace-RNN (V)  & 0.379  & 0.481 \\
			VGGFace-RNN (A+V) & 0.344 & 0.506 \\
			NISL (V)  \cite{deng2020multitask}  & 0.373	 & 0.513 \\
			$M^3$T (A+V) \cite{zhang2020m3t} & 0.320  & 0.550\\
			\hline
			Transformer (V)  & \textbf{0.381}  & $\textbf{ 0.489}^{\dagger}$ \\
			Transformer (A+V)  & $\textbf{0.396}^{\dagger}$  & $\textbf{0.525}^{\dagger}$\\
			\hline
			\hline
		\end{tabular}
	{\scriptsize $^{\dagger}$ indicates significant differences over corresponding VGGFace-RNN models.}
	\end{table}

	We first analyze the performance of the baseline models (Section \ref{ssec:baseline}). The AffWildNet models were trained purely on the video modality. Results illustrate that increasing the number of FC layers from 2000 to 4096 improves performance. Further, dynamically utilizing cues with the addition of recurrent layers gives 7\% absolute improvement for Valence and 6\% absolute improvement for Arousal.  The VGGFace-RNN(V) model retrained on the Aff-Wild2 shows an improvement of 12.4\% absolute for arousal and 6.4\% absolute for valence over the AffWildNet model.  
	The baseline for the audio modality, RNN(A) is inferior compared to the video only modality and has been noticed in other studies as well \cite{zhang2020m3t}. This maybe because in the Aff-Wild2 database, expressions are predominantly conveyed through the visual modality in the form of facial expressions and to a lesser extent through audio cues. A combination of the audio and visual modalities VGGFace-RNN (A+V) improves arousal prediction by 2.5\% whereas the valence prediction diminishes by 3.5\%. These results, similar to previous studies, indicate that predicting arousal values with audio cues is easier compared to valence values, as the tone of voice distinguishes arousal scores whereas visual features are more descriptive of the valence scores \cite{Sridhar_2018}. Among the {\color{black}external} studies, the NISL model shows slight improvement for arousal but not valence than other baselines. Similarly, $M^3T$ shows significantly better results for arousal but significantly worse for valence. Overall, these scores establish competitive baselines.
	
	Next we observe the performance of the proposed transformer model for predicting expression values. Table \ref{tab:baseline_results} illustrates gains for the proposed model over the VGGFace-RNN model for both the video (V) and audio-visual (A+V) modality. For the video modality we see gains of 0.2\% and 0.8\% for valence and arousal respectively. The addition of the audio cues further improve the performance of transformer models by 1.5\% and 3.6\% for valence and arousal respectively over using only video cues, indicating the benefits from attending across modality. Overall the audio-visual transformer model shows absolute gains of 1.7\% for valence and 1.9\% compared to the VGGFace-RNN model. {\color{black} Note that the proposed transformer models and our baseline VGGFace-RNN models have a similar number of parameters with less than 10\% difference between them.} {\color{black} Compared to the state of the art results from previous studies, we observe that the proposed models provide competitive results with significant improvements for valence and better in one case for arousal. Note that these previous studies were trained on extra data as well as a multi-task framework. The proposed models will potentially gain with these improvements.}

	\subsection{Ablation results}
	\label{ssec:ablation}
	\begin{figure}[tb]
		\centering
		\includegraphics[width=\linewidth]{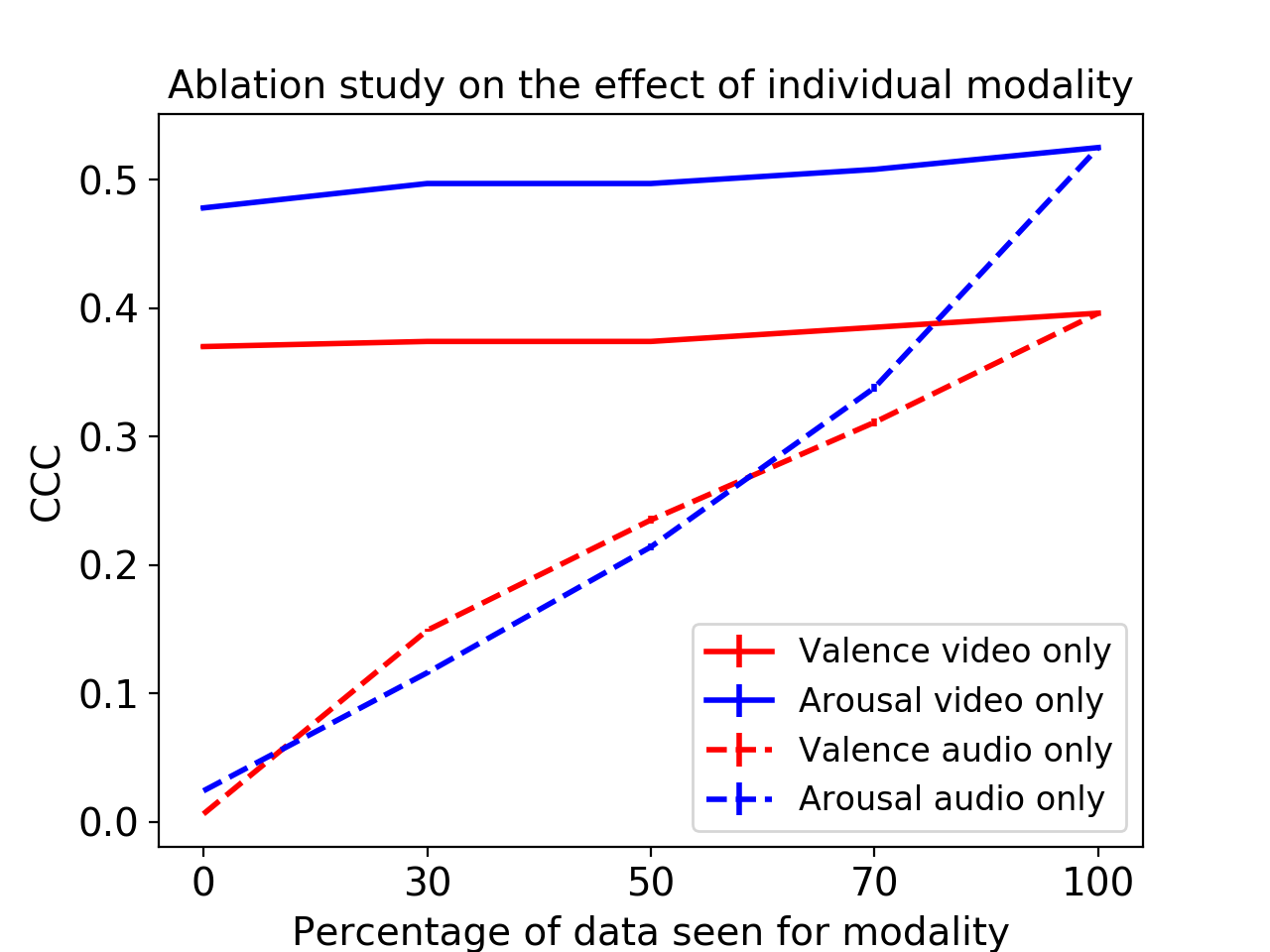}
		\caption{Ablation studies on the effect of individual modalities. Missing inputs are replaced by zeros.}
		\label{fig:ablation_results}
	\end{figure}
	
	We also conducted an ablation study to investigate the performance of the proposed model when either the audio or video modalities were missing. This study depicts a realistic scenario when the person moves away from the field of vision of the device they interact with or are not audible. We conducted the study by masking the inputs for the missing modality with zeros. The proportion of data missing is varied between 0 (modality completely missing) and 100 (modality completely present) to evaluate scenarios where the input is partially missing. The data to be masked is chosen randomly and we repeat the experiments over 10 trials on the Test Set. 
	
	Figure \ref{fig:ablation_results} illustrates the results. We observe that for both valence and arousal the complete absence of the video modality (audio only) severely degrades the performance producing CCC values close to 0. The model is unable to adapt when the visual cues are absent. Performance linearly improves with proportion of visual data seen.  On the other hand the complete absence of the audio modality (video only) slightly diminishes performance by approximately 2\%, producing results similar to the transformers trained on the video modality only. This shows adaptability of the audio-visual transformer architecture when audio data is missing. Overall, these results illustrate that the expression detection task is dependent on the rich visual information present in the Aff-Wild2 database.

	\section{Conclusion}

	This paper presented a multimodal architecture for expression detection on the Aff-Wild2 database. We utilized the transformer architecture with the addition of a cross-modal attention layer that attends to significant cues in the audio and video modality. A simple late fusion of the cues was then used to predict arousal and valence score for each video frame in the database. Comparison with competitive baselines and previous studies illustrate that the proposed models achieve state of the art performance for the expression detection task. Further the investigation of the contribution of each modality illustrated that the visual modality provides rich cues for the models to train on.

	Our future work will investigate the use of the multi-modal transformer for detecting expression categories. To further tap into the visual cues, we will study the use of landmarks as additional inputs for the models to train on. Along these lines we will analyze various methods to training models on missing data. We will also broaden the analysis of the proposed models on different data with additional modalities such as text.
	
	\bibliographystyle{IEEEbib}
	
	\bibliography{strings,reference}

\end{document}